Interferometric Space Missions for the Search for Terrestrial Exoplanets: Requirements on the Rejection Ratio

L Kaltenegger [1, 2], M. Fridlund [2], A. Karlsson [2]

[1] *Harvard-Smithsonian Center for Astrophysics, 60 Garden St, Cambridge, MA 02138, USA*
*e-mail: lkaltenegger@cfa.harvard.edu*
[2] *Research and Scientific Support Department, European Space Agency, ESTEC, Noordwijk, The Netherlands*
*e-mail:mfridlun@rssd.esa.int, akarlsson@rssd.esa.int*

**ABSTRACT**

The requirements on space missions designed to study Terrestrial exoplanets are discussed. We then investigate whether the design of such a mission, specifically the Darwin nulling interferometer, can be carried out in a simplified scenario. The key element here is accepting somewhat higher levels of stellar leakage. We establish detailed requirements resulting from the scientific rationale for the mission, and calculate detailed parameters for the stellar suppression required to achieve those requirements. We do this utilizing the Darwin input catalogue. The dominating noise source for most targets in this sample is essentially constant for all targets, while the leakage diminishes with the square of the distance. This means that the stellar leakage has an effect on the integration time only for the nearby stars, while for the more distant targets its influence decreases significantly. We assess the impact of different array configurations and nulling profiles and identify the stars for which the detection efficiency can be maximized.

## 1. INTRODUCTION

The direct detection of Earth-like exo-planets orbiting nearby stars, and the characterization of such planets - particularly as what concerns their evolution, their atmospheres and their ability to host life as we know it is one of mankind's great questions.

It is also a clearly defined scientific goal. Planets, which could host life, most likely will orbit within the so-called Habitable Zone (HZ), relatively close to the parental star. Further, as we understand it, these bodies must be rocky worlds, not unlike the Earth, Venus or Mars in our own Solar System, or at the least icy moons akin to Jupiter's larger satellites. The question of how large or how small such a world can be and still host life is one very likely demanding empirical data. To obtain such data, and to study the prevalence of Terrestrial planets, as well as their properties and ability to host life as we know it, now constitutes a high priority objective (theme) in the long-term science plan (Cosmic Vision 2015-2025) of the European Space Agency (ESA), as well as in the same of the US National Aeronautics and Space Administration (NASA). The goal in Cosmic Vision is to place our Solar System into context.

These agencies are currently working on the definition of instruments, carried on space missions, and capable of meeting the challenges posed by these goals. These missions, are respectively designated Darwin (Legér et al., 1996a, 1996b, Fridlund, 2000) and the Terrestrial Planet Finder (TPF – Beichman, 2000).

The direct detection of a planet, of the size of the Earth, orbiting its parent star in the HZ constitutes a challenging problem, since the signal detected from the planet is between about $10^{10-11}$ (visual wavelength range) and $10^{6-7}$ (mid-IR spectral range), times fainter than the signal received from the nearby star. Selecting the appropriate spectral region in which to attempt detection is governed by this contrast problem, *and* the selection of a region in which the characterization of the planet and its habitability is optimum. This problem has recently been addressed by a number of researchers e.g. Selsis (2002) and Traub (2003). The European mission Darwin has selected the spectral region between 6μm and 20μm, a region that contain (among others) the $CO_2$, $H_2O$, $CH_4$, and the $O_3$ spectral features found in the terrestrial atmosphere. The presence or absence of these spectral features would indicate similarities or differences with respect to the atmospheres of known telluric planets such as Venus, Earth and Mars.

A fundamental part of the problem of *directly* detecting the planet with its feeble light in the glare of the strong parental stellar flux, thus is the huge contrast. Several techniques have been suggested, most notably giant (100m class) ground based telescopes operating in the visual or near-IR, large (10m class) space deployed coronographic telescopes operating in the visual or interferometers operating in the visual or mid-IR. The selection is a complex issue, described in another paper (Fridlund et al., 2006), but including elements such as compatibility with the scientific case, the difficulty of implementation, the time schedules and costs. ESA has selected an Interferometer operating in the mid-IR and utilizing free flying telescopes (i.e. no connected structures) for detailed study and possible implementation. This is what is commonly referred to as the Darwin mission (Fridlund, 2000), which further is using the new technology of nulling (or destructive on-axis) interferometry (Bracewell, 1978; Bracewell & McPhee, 1979). The basic concept here is to sample the incoming wavefront from the star and its planet(s) with several ($\geq 2$) telescopes that individually do not resolve the system. By applying the suitable phase shifts between different telescopes in this interferometer array, destructive interference is achieved on the optical axis of the system in the combined beam. At the same time, constructive interference is realized a short distance away from the optical axis. Through the appropriate choice of configurations and distances, one can for the specific case, place areas of constructive interference on regions representative of the HZ and so achieve the required contrast. The first practical demonstration of nulling on the ground was undertaken in February 1998 (Hinz et al., 1998). Using the Multiple Mirror telescope on Mount Hopkins, Arizona, these authors were able to cancel out the image of a star: α-Orionis. The ability of the interferometer to suppress the entire Airy pattern was demonstrated and in this case the nulled image had a peak in intensity of 4.0% and a total integrated flux of 6.0% of the constructive image.

Planet finding missions like the Darwin mission or the Terrestrial Planet Finder could be implemented in a wide variety of different nulling interferometer architectures (e.g. Karlsson et al., 2004, Coulter, 2003) and configurations, constrained by the number of telescopes and the necessary background and starlight suppression. Recent work suggests that requirements on the shape of the null can be relaxed when one takes into account unavoidable noise contributions introduced by instrumental errors (Kaltenegger & Karlsson, 2004; Dubovitsky & Lay, 2004). Thus configurations with fewer telescopes can be investigated as candidates for the Darwin mission reducing complexity and cost of the mission.

In this paper, we will first briefly specify and discuss the scientific requirements on such a mission (for more detail, see Fridlund et al., 2006). This is followed by a brief introduction to aspects of nulling interferometry. We then discuss the background signals and noise terms and how they contribute to the signal to noise calculations. Finally, we discuss the issues of different configurations and number of telescopes and how they affect the fulfillment of the high level scientific requirements when confronted with the actual stellar sample to be observed.

## 2. THE HIGH LEVEL SCIENTIFIC REQUIREMENTS

Due to the cost and complexity of any space mission, it is paramount that the scientific objectives of the mission are specified in a stringent way. This allows the transformation of these goals into a set of requirements for the mission that can be converted into a design. It is of course an iterative process, eventually coalescing into fixed parameters. In the case of the terrestrial exoplanet mission of ESA (Darwin) this is based on a proposal (Léger et al., 1996a, 1996b). The ideas in this work in its turn built on earlier theoretical work (Bracewell & McPhee, 1979; Angel et al., 1986). This proposal led ESA to carry out a feasibility study between 1997 and 2000, resulting in the principles governing the project today (Fridlund, 2000). The overarching goal, as formulated in this study is specified as: *To detect and study Earth-type planets and characterize them as possible abodes of life.*

In Fridlund (2000), the high level scientific requirements were described as being the answer to the following questions:

- Are we alone in the Universe?
- How unique is the Earth as a planet?
- How unique is life in the Universe?

The feasibility study carried out by ESA in order to investigate if it was currently technically possible to address these questions led to – after a number of detailed technical studies – what is currently being represented by the name Darwin and is being designed to establish the following:

1. Search nearby stars for Terrestrial planets
2. Detect planets within the so-called 'Habitable Zone', i.e. the planetary orbital radii around a specific star where water could be found in a liquid state. In the Darwin study, the HZ was considered only in terms of a black body temperature. No provision was made to take into account atmospheric pressure, etc

3. Determine the planets orbital characteristics (period, eccentricity, inclination etc.)
4. Observe the spectrum of the planet. Detection of the presence of an atmosphere, effective temperature, diameter of planet. Determine the composition of the atmosphere, viz. the presence of water, ozone/oxygen in an Earth type planet, mainly inert gases in a Mars/Venus type planet and Hydrogen/Methane atmospheres in Jupiter type planets or 'primordial' Earth-like planets

These high level scientific requirements can be translated into specific observational requirements that can be translated into mission requirements. These requirements are detailed in Fridlund et al (2006), but can be summarized briefly as:

1. Minimum number of single, solar type (F-K main sequence) stars to be surveyed for Terrestrial exoplanets in the HZ during primary mission – 165 to 500. M-dwarfs, the most commonest stars in the Galaxy are also considered.
    a. 165 under the added condition that significantly amounts of dust (10 times level in solar system) is present (see below)
    b. 500 (complete sample of F, G and K single stars out to 25pc + a large number of M dwarfs) under the conditions of similar levels of dust to the solar system
2. Completeness of survey (probability that one has not missed a planet in the HZ for a specific star) – 90%
3. Spectral signatures to be observed for each planet in a detected system – $CO_2$, $H_2O$, $CH_4$, and $O_3$

Detailed specifications can be found in the abovementioned paper (Fridlund et al., 2006). Technically, Darwin is a nulling interferometry using modulation, which being so far not implemented in any operational observatory at those wavelengths and with the complexity required in space. Darwin therefore will require significant development before a mission can be launched.

Recently, the scientific rationale given for the Darwin mission has been put into the broader context of ESA's new science plan for the period 2015 – 2025. This plan, designated Cosmic Vision, covers the major questions to be addressed by European space science in 4 themes, the first of which is "What are the conditions for Planet formation and the emergence of life?" Darwin is an integral element within this theme, addressing the finding of Terrestrial planets around nearby stars, as well as a first determination of their physical parameters – including their habitability.

### 3. NULLING INTERFEROMETRY, CONFIGURATIONS AND THE FORMULATION OF THE PROBLEM

The simplest nulling interferometer is the so-called Bracewell array (Bracewell, 1978), which consists of two telescopes whose output combines with a π phase shift. The beams from each of the telescopes are made to interfere destructively (the process known as nulling) on the optical axis. Thus light from a star placed on the optical axis of each of the telescopes and thus of the interferometer can be centered on a deep null, and effectively extinguished. Light from a point source, separated by a small angular distance, θ, will arrive with a small external phase delay, θ B cos(φ/λ), where B is the distance between the two telescopes and θ is the azimuthal sky coordinate of the source.

In the absence of perturbations, the interferometric transmission of the Bracewell interferometer with two circular entrance apertures of diameter D for a point source at sky coordinates (θ,φ) is given by equation 1:

$$T_{Bw}(\theta,\phi,\lambda) = 2 \times \frac{J_1(\pi\theta D/\lambda)}{(\pi\theta D/\lambda)^2} \sin^2\left(\pi \frac{\theta B}{\lambda} \cos(\phi)\right), \quad (1)$$

where $J_1$ is the Bessel function. Equation 1 shows that the interferometer response is (for small angles) proportional to $(\theta B/\lambda)^2$. Because of the shape of the destructive interference fringe, as a function of position, we refer to this as a $\theta^2$ null. Configurations based on a number of Bracewell pairs, arranged in certain patterns will have destructive responses with a $\theta^4$ or $\theta^6$ (a wider) null and thus a better suppression of excursions from the on-axis flux (of e.g. a stellar disk) and thus a lower stellar leakage. Thus, while an ideal interferometer with a π phase shift would have a 100 % null at an infinitesimal bandwidth, in reality, the nulling 'band' will have a thickness, and a shape depending on B, θ and the configuration (including the number of telescopes making up the arrangement). Due to the size of the stellar disk a residual stellar signal $S_{leak}$ will pass through the central transmission pattern. This leak is calculated according to equation 2, and the shape of the null for two different configurations (representative of $\theta^2$ and $\theta^4$ null profiles) can be seen in Fig. 1 and Fig. 2. The sensitivity pattern plotted onto the sky is designated the transmission map, and is most easily visualized by imagining the transmission pattern in Figure 2 to be rotated by 2π out of the plane of the paper.

$$S_{leak}(\lambda) = A_{tot} \int_\theta \int_\phi T(\lambda,\theta,\phi) B_{star} d\theta d\phi \quad (2)$$

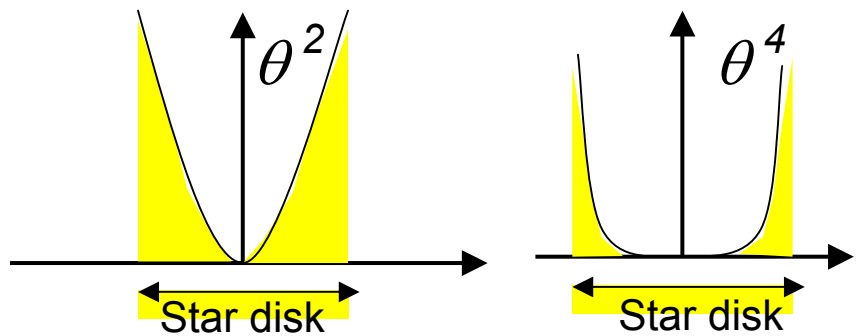

Figure 1 Different shape of the central null and rejection of light from the stellar disk for the $\theta^2$ and $\theta^4$ configurations

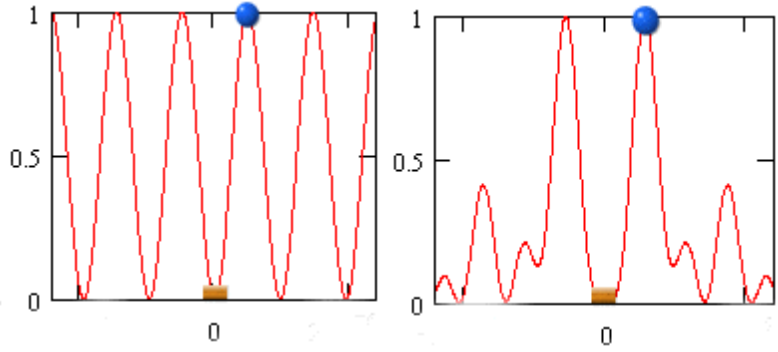

Figure 2 Transmission pattern of a configuration with a $\theta^2$ ( Bracewell, 1978) and a $\theta^4$ ( the Bowtie configuration as suggested by Absil, 2001) geometry null. A radius is displayed from the center (photocenter of the star) out to a radius of 5AU. Both cases are assumed observed at a distance from the Earth of 10pc. The stellar disk is magnified and a planet is shown at 1AU (blue dot).

In practice the central null of the transmission map will be degraded by amplitude and optical path differences between the interferometer arms that will lead to a level of noise that sets a limit for the achievable null depth given (Serabyn, 2000). Averaging over the detector readout time leads to equation (3).

$$\frac{\langle F_{leak} \rangle}{F_s} \approx \frac{\alpha^2}{4} + \frac{4\Delta\phi \cdot \alpha}{3\pi} + \delta_A^2 + \frac{\Delta\phi^2}{4} \qquad (3)$$

Here $\alpha = \pi\theta_S B / \lambda$. The $\Delta\phi$ term describe the wavelength dependent dispersion effects while $\sigma_A$ is the root mean square fluctuation of the relative amplitude between the two arms of the interferometer after spatial filtering. The leakage term and the resulting null degradation depends strongly on the baseline used. Optimizing the baseline of the interferometer for each stellar target system will minimize the leakage, while a fixed baseline as would be used for an interferometer implemented on a structure would lead to a high level of leakage for most of the target systems. In the input catalogue for searches around nearby stars, distances vary between 3pc and 25pc (Kaltenegger et al., 2006). It would therefore be necessary, in the case of a structurally mounted interferometer, to implement a mechanism that allow for a variation of optimized baselines. Clearly this optimization of the leakage strongly supports an interferometer deployed on free flying spacecrafts.

The stellar leakage is, however, not the dominant noise source for stars with distances larger than about 10pc. Instead, local and exo-zodi will dominate. This has consequences for the selection of interferometric architecture, since only the integration time for the closest stars in the input catalogue would be negatively affected by a reduction of the null width from a $\theta^4$ to $\theta^2$ profile (Dubovitsky & Lay, 2004; Kaltenegger & Karlsson, 2004; Karlsson et al., 2004). Since the number of single, accessible solar type stars within 10pc is relatively small, a total of 12 stars excluding 46 M-dwarfs of the in total more than 1135 candidates in the target star catalogue out to 25pc. Of these 628 are considered single and 241 of the latter are classified as M-dwarfs (Kaltenegger et al., 2006). The majority of targets are thus found at a distance where the local zodiacal cloud instead is the dominant noise factor (section 5). Our detailed calculation (section 6) have independently shown that one can relax the starlight rejection criteria of the nulling interferometer mission while still observe a target star sample of a required minimum of 150 to 500 stars within the allotted mission primary life time, thus meeting the mission requirements. The original Darwin mission concept (Fridlund 2000) minimized stellar rejection in order to optimize the search for Earth-like planets orbiting the very closest target stars. This led to a baseline configuration with 6 free flying telescope spacecraft and a central beam combiner, as originally suggested by Jean-Marie Mariotti, and later modified and developed into the Bowtie configuration (Absil, 2002). The fact that we are now considering configurations optimized for the much larger sample of stars at distances greater than 10pc does not mean that the nearby stars have been discarded. Since in the case of the nearby objects the stellar leakage is greater it will take a longer time, however, to acquire the required signal-to-noise. The planets will have an intrinsically higher flux, which in most cases will compensate for the increased noise.

Because of this realization, recent candidate Darwin and TPF configurations tend to use three or four telescopes (e.g. Karlsson et al., 2004; Coulter, 2003) significantly reducing complexity and cost of the mission. Three telescopes is the minimum number of telescopes needed for an interferometer mission that uses rapid signal modulation with sub-interferometers (Absil et al., 2002) to detect a planet in the high background noise.

The use of a single Bracewell as a nulling interferometer for planet detection is limited by the difficulty to rapidly modulate the planet signal against disturbances such as stellar leakage and the background. For a system at 10pc observed at 10 µm the local zodiacal cloud signal will be the biggest noise source. Second in importance is the stellar leakage followed by the signal of the exo-zodiacal dust disk (if present). Here we have assumed an exo-zodiacal dust disk similar to our local zodiacal cloud. The planetary signal will be hidden in that background and fast modulation is required in order to detect a potential planet. Modulation could be achieved by rotation of an interferometric array. The problem with such an approach will be that such a modulation will be necessarily slow because of limitations in operating spacecraft. The modulated planet signal could thus easily be corrupted by introducing drifts leading to *1/f* noise. An alternative modulation technique is internal modulation, which is based on combining the outputs of various sub-arrays e.g. two Bracewell arrays, and implementing a variable phase shift (Absil, Karlsson and Kaltenegger, 2002). Alternative modulation techniques are under investigation especially in the context of e.g GENIE, a ground-based pre-cursor for Darwin planned for implementation at the European Southern Observatory's Very Large Telescope Interferometer (VLTI) in Chile (see e.g. Gondoin et al., 2004) and smaller space based interferometric arrays carrying our technological verification and science.

## 4. BACKGROUND AND FOREGROUND FLUX

The detected signal to noise is governed by the contribution of separate background and foreground noise terms to the interferometer output signal. These noise terms can be divided into instrumental noise, such as thermal background from the telescopes and their components, and other constant and variable noise sources caused by imperfections in the system, and noise sources caused by other astronomical objects. Excluding the possibility of a strong background (e.g. a galaxy) or foreground (e.g. a solar system object) IR source falling within the field of view, these noise sources consist mainly of the local solar system zodiacal light caused by interplanetary dust and exozodiacal light caused by similar dust in the target systems. The local zodiacal cloud (LZ) provides the foreground through which Darwin will have to observe. Because this zodiacal foreground is diffuse, it cannot be cancelled by nulling interferometry. Thermal emission of the exo-zodiacal (EZ) dust will be the strongest source of noise photons at short wavelengths. Currently, the range of EZ expected from target systems is unknown. It is only very recently, that observations with NASA's SPITZER infrared observatory have detected Kuiper–Edgeworth-belt (K-E belt) dust around normal main sequence stars (e.g. Kim et al., 2005; Bryden et al, 2006). This dust, although similar in composition is located much farther out both in the Solar system and in the target systems. In the solar system the K-E belt dust originate outside Saturn where it is presumably created by collisions between the aggregate of bodies orbiting there. Kept into place temporarily by resonances, it will on timescales of thousands of years disappear through either Poynting-Robertson effect or through photon pressure. Any observed dust must thus be continuously replenished, and would imply a significant repository of cometary or asteroidal bodies. In one case (so far) SPITZER see warm dust – which thus originate at most a few AU from the central star – and no cold (K-E belt) dust around a late K-type star (Beichman et al., 2005). In this case the amount of dust is in excess of 1000 times that found in our solar system.

The level of EZ dust will determine the efficiency with which the missions like Darwin will be able to discern an Earth-like planet. As we do not know in what fraction of stars the dust levels will be prohibitively high, for our calculations we have to assume a range within which it will be possible to detect terrestrial planets. Here our assumption is a 'standard case' of 1 EZ (being equal to the Solar System value), and a 'worst case scenario' of 10 exo-zodies. It would be possible, in principle to detect the Earth against a background of 30 exo-zodies, but this would come at the price of very long integration times

The LZ and EZ diminish with distance from the Sun and the star respectively, due to decreased temperature and decreased dust density. Note that large coherent structures such as wakes and clumps behind planets can masquerade as planets. Those structures could also serve as markers for the presence of planets, if their location with respect to a planet were well understood. Exo-zodiacal clouds are thus not expected to be uniform, and a planet must be detected against a non-flat field of corrugations. In our own cloud, these 'clumps' are assumed to have roughly less than 0.1% of the amplitude of the total cloud brightness and are thus no source of confusion.

The stellar leakage is dominating for short wavelengths, while the thermal background level due to the emission of the optics, can approach that of the zodiacal dust for temperatures above 40K at the longer wavelengths. However, the LZ dust disk remains the biggest noise factor, for most of the wavelength band.

The analytical model used for calculating the EZ in our calculation is building on the one by Kelsall et al. (1998) describing our own LZ cloud. This model is based on observations of the solar system made by the Diffuse InfraRed Background Experiment (DIRBE) aboard the COsmic Background Experiment (COBE) satellite. The zodiacal cloud in our Solar System extends outward at least to the asteroid belt, roughly 3.5 AU, and inwards to the Solar corona, at a few solar radii. As an inner cutoff for the exo-zodiacal cloud we use a sublimation temperature of 1500K. The outer cutoff is given by the field of view of the interferometer which is limited by the acceptance angle of the fiber used to clean the wavefront (see above).

$$B_{EZ} = \int B(T(r))\rho_0 r^{-\alpha} e^{-\beta\left(\frac{z}{r}\right)^{\gamma}} dl \qquad (10)$$

with $\alpha = 1.39$, $\beta = 3.26$, $\gamma = 1.02$, $\rho_0 = 1.14\ 10^{-7}$,

$$T(r) = T_0 r^{-\delta} \left(\frac{L}{L_{sun}}\right)^{\xi} \qquad (11)$$

with $\delta = 0.42$, $T_0 = 286$, $\xi = 0.234$

The flux of the local zodiacal dust disk was fit to a model also based on the same model by Kelsall et al. (1998). Note that there is a small difference in the local zodiacal flux due to viewing direction to the ecliptic latitude of the observed star systems. Our computations takes this into account.

## 5. THE MODEL

For a practical observational scenario, we here assume a target system consisting of one or multiple planets, an exo-zodiacal dust disk and a host star. Obviously also the local zodiacal dust disk and the thermal radiation of the instrument at 40K have to be accounted for.

In general, the nulled output signal $F(t)$ for an extended source of brightness distribution $B_{disk}(\theta,\phi)$ that is transmitted through the interferometer pattern, can be determined using equation 4.

$$F(t) = A_{tot} \int_\lambda \int_\phi \int_\theta B_{Disk}(\theta,\phi)\ T(\theta,\phi,\lambda,t)\ \theta\, d\theta\, d\phi\, d\lambda, \qquad (4)$$

where $A_{tot}$ is the total collecting area of the interferometer, and T the normalized response of the interferometer, the transmission. The instrumental transmission is wavelength dependent and therefore the baseline of the array has to be optimized such that the response at the assumed planet position, $T_{planet}(\lambda,t)$, is maximized over an as wide part of the wavelength range as possible. In an alternative scenario the array can be reconfigured during the observation in order to provide a high transmission of the planetary signal over several selected wavebands in turn. When sub-interferometers are used to modulate the planetary signal, the overall interferometer response is called the Modulation $M(\theta,\phi,\lambda,t)$ and can be used instead of the $T(\theta,\phi,\lambda,t)$ in the calculations.

We model the star and the planet as uniform disks of angular radius $\theta_S$ and $\theta_{planet}$ respectively and use Planck's equation to obtain the brightness distribution $B_{star}$ and $B_{planet}$ as a function of wavelength. $T_{planet}(\lambda,t)$ denotes the response of the array at the planet's position over wavelength and time with a baseline optimized for its detection. When the planet search is performed by rotating the array, the mean value of the transmission value over that orbit can be used in a first approximation to determine the signal strength. The wavelength dependence of the interferometer's response allows extraction of basic color information of the detected planet because the response of the interferometer is well known. This technique could be used to rapidly distinguish giant planets from terrestrial planets during the detection phase of the mission. We thus have the following set of equations contributing to our model.

$$F_{planet}(t) = A_{tot} \int_\lambda B_{planet}(\lambda)\ T_{planet}(\lambda,t)\, \pi \theta_{planet}^2\, d\lambda, \tag{5}$$

$$F_{leak}(t) = A_{tot} \int_\lambda \int_\phi \int_\theta B_{Star}(\theta,\phi,\lambda)\ T(\theta,\phi,\lambda,t)\ \theta\, d\theta\, d\phi\, d\lambda + \langle F_{leak} \rangle \tag{6}$$

$$F_{EZ}(t) = A_{tot} \int_\lambda \int_\phi \int_{FOV} B_{EZ}(\theta,\phi,\lambda)\ T(\theta,\phi,\lambda,t)\ \theta\, d\theta\, d\phi\, d\lambda \tag{7}$$

$$F_{LZ}(t) = A_{tot} \int_\lambda B_{LZ}(\theta,\phi,\lambda)\ \int_\phi\!\!\int_{S\Omega} T(\theta,\phi,\lambda,t) \theta\, d\theta\, d\phi\, d\lambda, \tag{8}$$

$$F_{thermal}(t) = A_{tot} \int_\lambda B_{thermal}(\theta,\phi,\lambda)\ \int_\phi\!\!\int_{S\Omega} T(\theta,\phi,\lambda,t) \theta\, d\theta\, d\phi\, d\lambda, \tag{9}$$

Only the flux of the exo-zodiacal dust disk $F_{EZ}$, the star $F_{leak}$ and the planet $F_{planet}$ are transmitted through the interferometer pattern. The flux from the local zodiacal dust disk $F_{LZ}$ as well as the thermal background flux, $F_{thermal}$ are not interfering destructively.
Spatial filtering of the high frequency wavefront errors by monomode fibers or waveguides is envisioned for Darwin (Karlsson et al., 2004). The geometric extent of the beam seen by such a fiber or waveguide is given by $S\Omega \approx \lambda^2$. The IR thermal background as well as LZ contribution is proportional to $S\Omega$ while the signals in the field of view of the interferometer like the planet and the EZ are proportional to the interferometer's collecting area. Note that the change of the coupling of the light from different off-axis angles into the fiber over the Field of View has been approximated by a constant factor in the calculations.

Note that the recently discovered 'variability noise' ( Lay & Dubovitsky, 2004) terms have not been included in our calculations. This omission does not change any of the conclusions in this paper with respect to configurations, etc.

## 6. MODEL CALCULATIONS

In Figure 3 and 4 we demonstrate the impact of various noise terms, as a function of wavelength, in our calculations. The term 'nullfloor' refer to the actual level of transmission in the center of the null. Due to the imperfections in the system some signal will always be carried through. This is not the same as the term 'stellar leakage' that refers to the signal passed through due to the resolved surface of the star (and the variations in this signal du to vibrations, etc). Even in the case of an ideal point source, some flux would be transmitted through the bottom of the null.

In the case of our model, we assume a constant phase shift offset between the two sub-arrays of $10^{-3}$, and a root mean square fluctuation in matching amplitudes of $10^{-2}$. Polarization and any other systematic error have been ignored for this calculation. The resulting graphs clearly demonstrate the influence of the distance to the observed star system on the background signals and accordingly on the integration time to detect and observe a planet assuming a $\theta^2$ configuration, i.e. a configuration utilizing 3 telescopes. If we assume a rejection ratio of at least $10^5$, the stellar leakage is most important for the shorter wavelengths. At the large wavelength end the thermal background level due to the emission of the optics can approach that of the zodiacal dust for temperatures above ~40K.

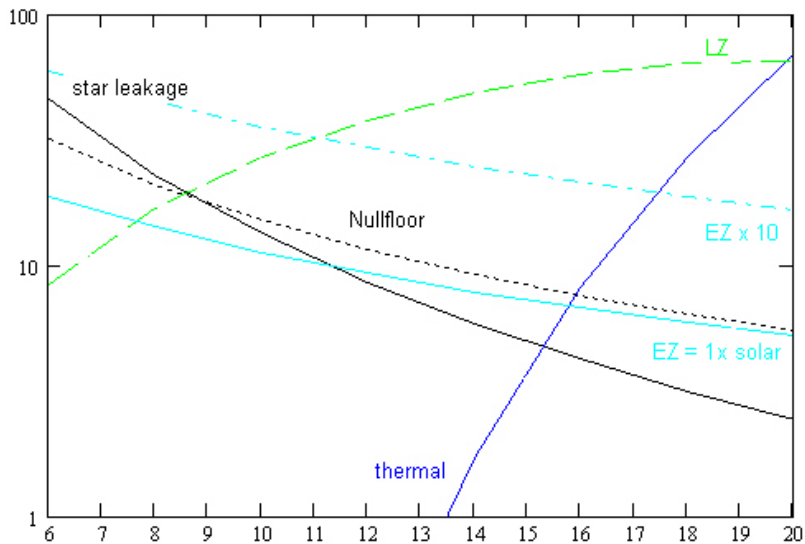

Figure 3: Comparison of noise terms for a G star system at 10pc as a function of the wavelength range intended for the Darwin mission. Note the result of increasing the exozodiacal dust level to 10 times that of the Solar System. The optical train is assumed to be at a temperature of 40K.

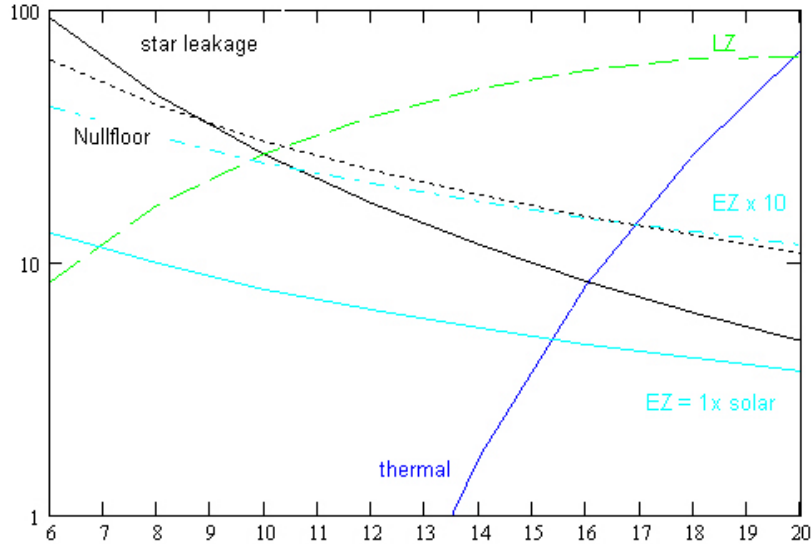

Figure 4: Comparison of noise terms for a G star system at 5pc as a function of the wavelength range intended for the Darwin mission. Note that the stellar leakage for short wavelengths now is higher than the exozodiacal dust level of 10 times that of the Solar System. The term Nullfloor refers to matching errors being passed through the central null (see text). The optical train is assumed to be at a temperature of 40K.

In an optimum scenario the planet will lie on the first interference maximum for a given wavelength or wavelength-band. Due to collision avoidance the Inter-Satellite Distance (ISD) in a free flying interferometric array will constrain the minimum possible baseline. This means that for the closest target stars the planet cannot be put on the primary maximum in the transmission pattern. Instead a secondary maximum or one of the following peaks must then be utilized. This is possible since the transmission and thus also the modulation pattern repeats symmetrically. It is noteworthy that this will increase the stellar leakage, as the extent of the central null naturally also scales with the baseline, as can be seen in Figure 5. For a close-by star (the example shown is at 5pc), the ISD sets the minimum baseline that can be used.

From these results, comparing with the other noise terms (e.g. Figures 3 & 4), it is clear that for G-stars at a distance further away than 10pc the integration time will not be governed by the stellar leakage but instead by the local zodiacal light, except for the very shortest wavelengths.

If one considers the minimum ISD for both $\theta^2$ and $\theta^4$ configurations the leakage generally increases for close targets stars as neither architecture can put the planet on the first maxima and simultaneously maintain the minimum ISD (assumed to be 10m in these calculations). The difference in stellar leakage between the $\theta^2$ and $\theta^4$ null decreases with the number of transmission peaks the planet is traversing. In previous architecture studies for nulling interferometry, the ISD has not been considered. Taking it into account lessens the advantage of a $\theta^4$ configuration over a $\theta^2$ configuration significantly.

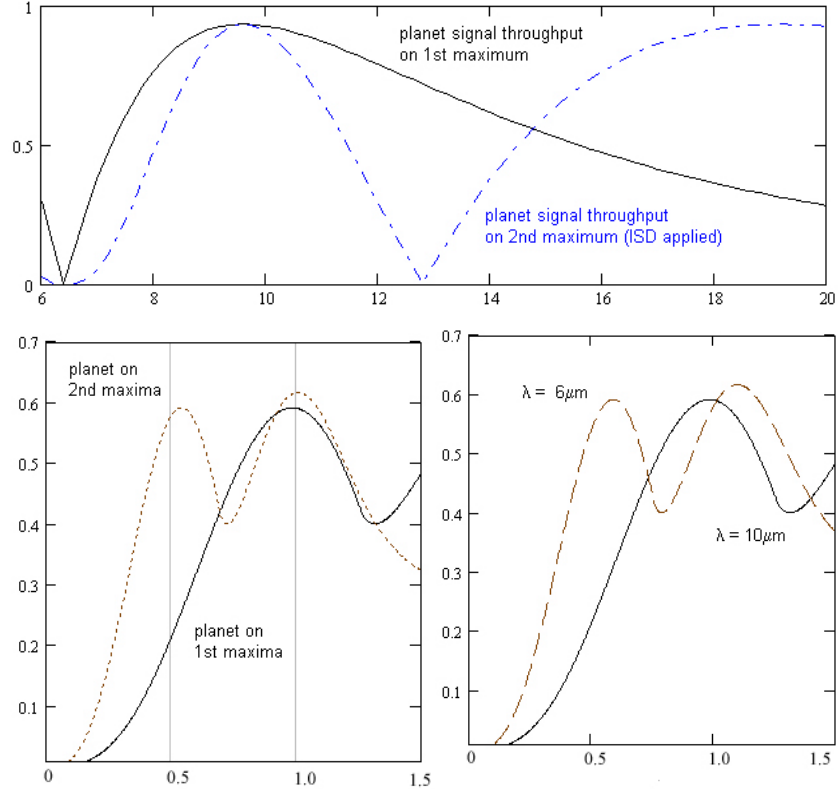

Figure 5 Transmission of the planetary signal (star planet system at a distance of 5pc) as a function of wavelength in one exposure (planet on 1$^{st}$ and 2nd transmission maximum) (upper panel), mean transmission after a full 360 degree rotation of the array (lower panel left, planet on 1$^{st}$ and 2nd transmission maximum), mean transmission of the planet signal after a full 360 degree rotation of the array for different wavelengths (lower panel (x-axis out to 1.5AU))

Some wavelengths will transmit extremely low levels of planetary light depending on the transmission pattern of the interferometer and the number of transmission peaks onto which the planet is placed. As an example, Figure 5 shows how the first and second transmission peaks influence the planet's signal transmission. For some configurations the rotation of the array can solve that problem by moving the planet into more sensitive areas (of the transmission pattern on the sky) while in other cases a resizing of the baselines will allow one to obtain spectral information over the complete band. This effect will introduce basic color information for the planet around close-by stars already in the detection phase where we have no a priori planetary information. This could be used to determine planetary properties already in the survey/discovery phase. In the spectroscopy phase of the mission the baseline can be optimized so that the planet flux is at a maximum at the observed wavelength channels.

## 7. CONSEQUENCES OF DIFFERENT CONFIGURATIONS ON THE SIGNAL TO NOISE RATIO

The signal-to-noise ratio (SNR) can be calculated accordingly:

$$SNR \approx \frac{T_{planet} F_{planet}}{\sqrt{F_{Background}}} \sqrt{t_{obs}} \qquad (12)$$

The increase in stellar leakage for a $\theta^2$ null configuration with respect to a $\theta^4$ null will lower the achievable SNR, most significantly for the nearby stars (a few to 10pc). On the other hand the maximum modulation efficiency of an array with a $\theta^2$ null shape will be higher, counteracting the influence of the increased background level. Especially at distances where the local zodiacal cloud is the major background noise, an increase in transmitted planetary photon flux will increase the SNR. One has to keep in mind that spectroscopic measurements at the lower end of the 6μm – 20μm waveband will require longer integration times. Additionally if the flux of the exo-zodiacal dust disk is higher than the assumed 1 local zodiacal level, the shape of the central null will become increasingly important since a wider null will block out a wider section of the hot (brighter) inner part of the EZ disk.

Figure 6 compares the stellar leakage for a 3 telescope TTN configuration (Karlsson et al., 2004) that has a $\theta^2$ null (solid line) versus the 6-telescope Bowtie configuration (Absil, 2001) providing a $\theta^4$ null (dashed line). Both performances are given as a ratio between the increase in the signal to noise and an ideal system with no leaks. For the detailed performance of the TTN architecture see Karlsson et al (2004).

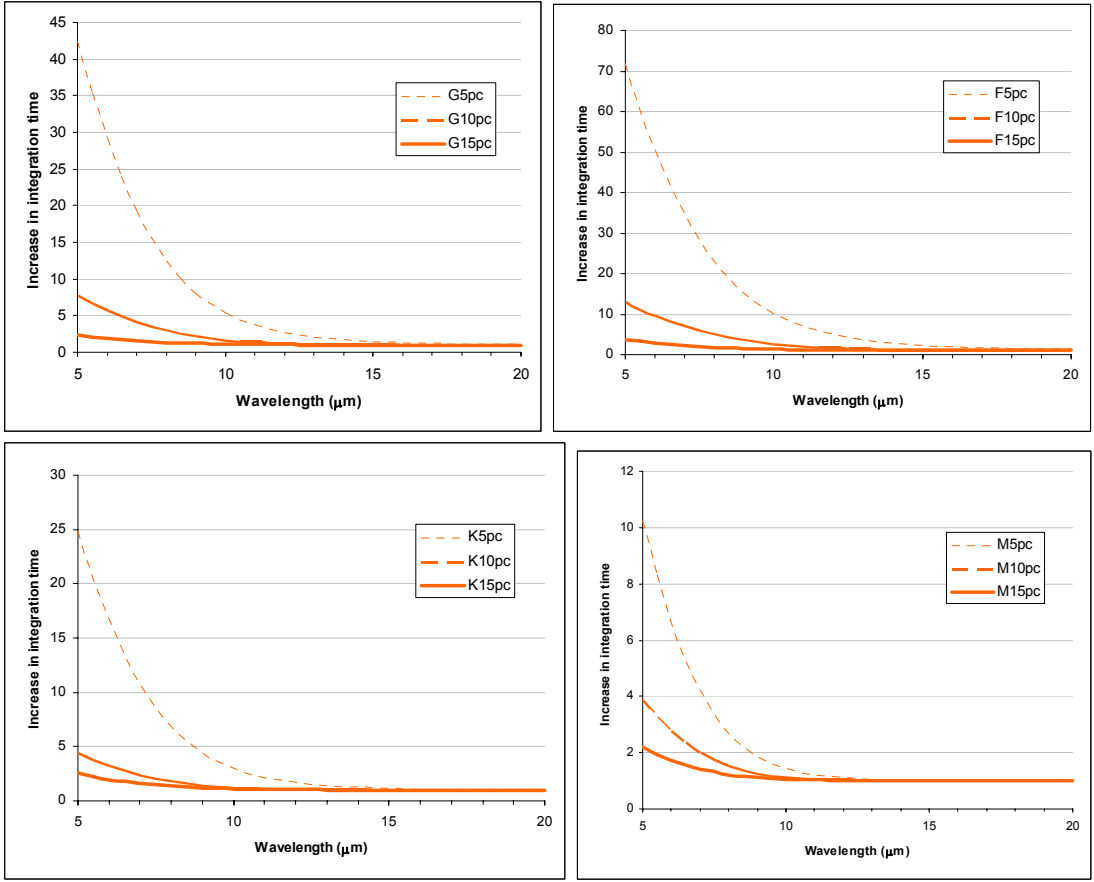

Figure 6: Increase in SNR per second for a TTN $\theta^2$ leakage for a F2V, G2V; K2V and M2V star at 15pc (solid line), 10pc (dashed line) and 5pc (pointed line) relative to an ideal non-leakage. Note the different y-scale.

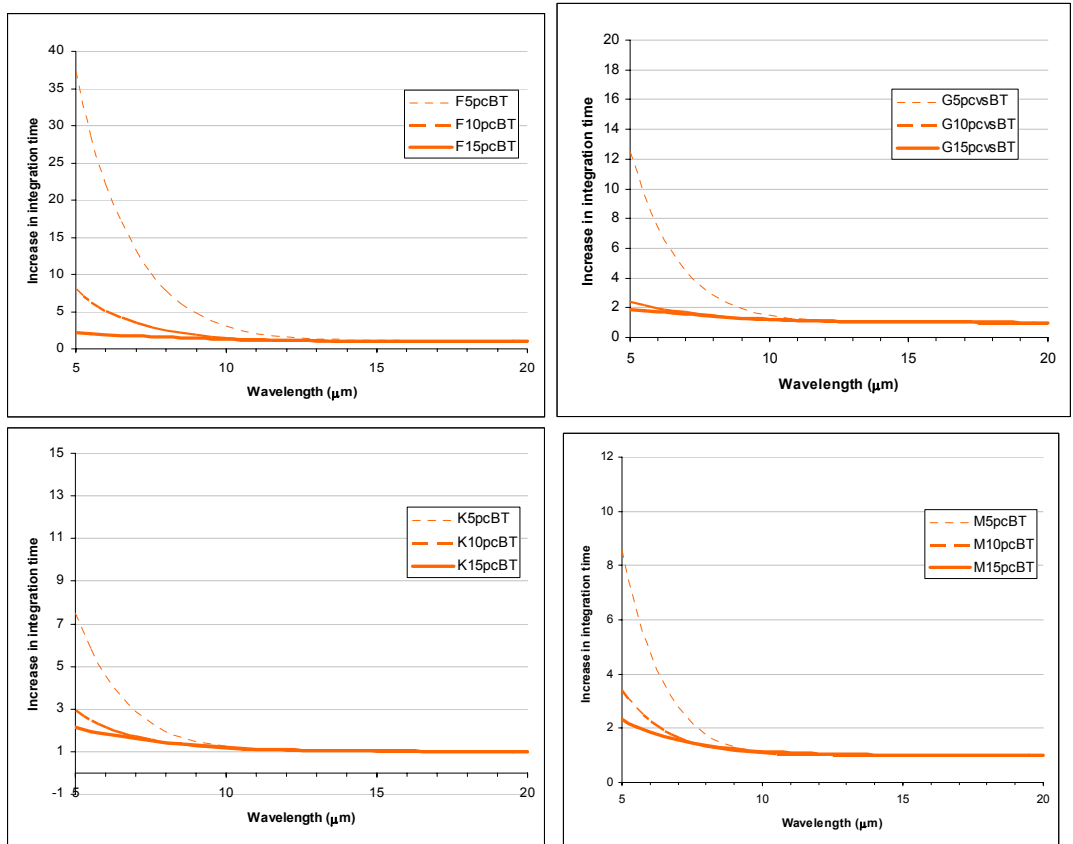

Figure 7: Increase in SNR per second for a Bowtie $\theta^4$ leakage relative to an ideal non-leakage calculation for a F2V, G2V; K2V and M2V star at 15pc (solid line), 10pc (dashed line) and 5pc (pointed line). Note the different y-scale.

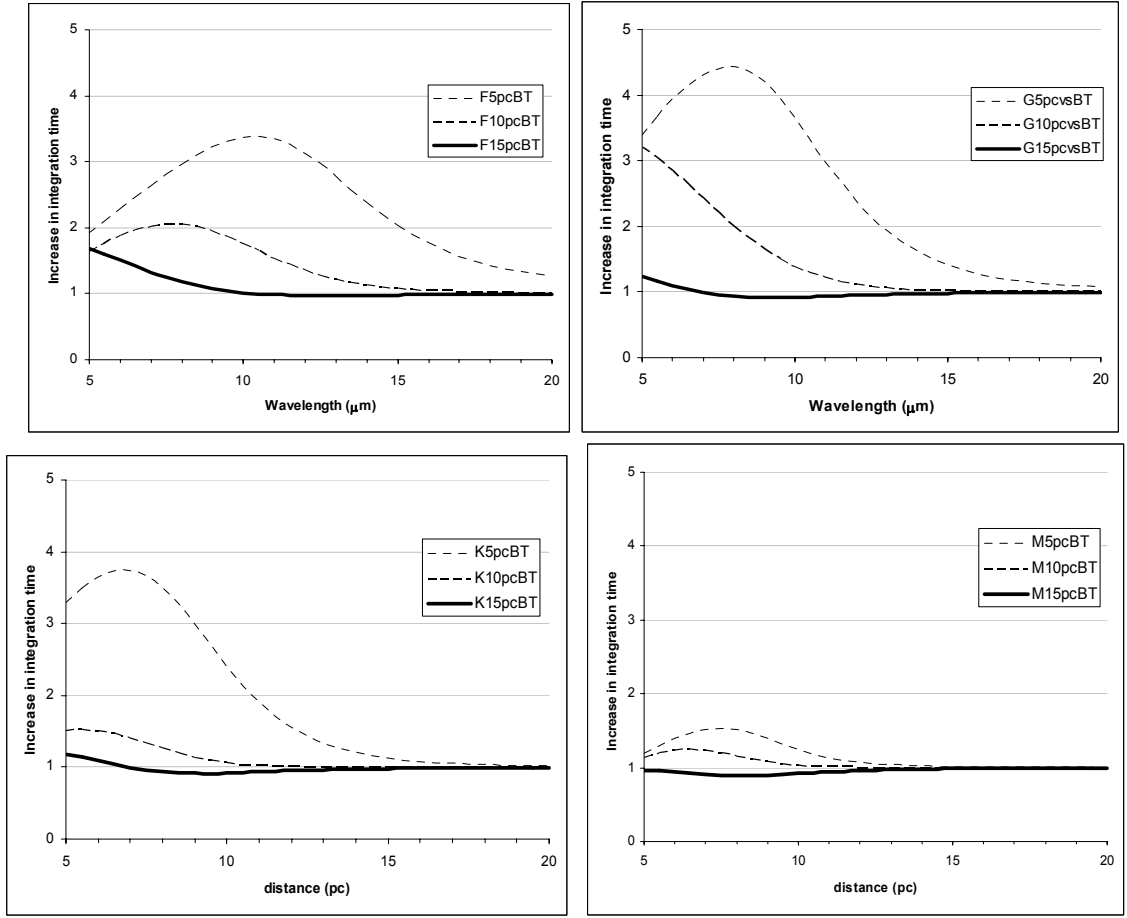

Figure 8: Increase in SNR per second for a TTN $\theta^2$ leakage (solid line) versus a Bowtie $\theta^4$ leakage for a F2V, G2V; K2V and M2V star at 15pc (solid line), 10pc (dashed line) and 5pc (pointed line).

Figure 8 shows that the increase in integration time is only relevant for the short wavelengths and stars at distances smaller than 10pc.

Most of the nearby target stars, selected for the search (e.g. Kaltenegger, 2004, Eiroa, 2003) are beyond that distance. For a consistent sample of more than several hundred F, G, K and M stars, the observations could in principle be carried out in the volume between a distance of 10pc and 25pc. If we chose, for instance, the scenario where Darwin can observe between +45 and –45 degrees ecliptic latitude (e.g. Fridlund, 2000), there are 29 F, 66 G and 160 K single stars between 10pc and 25pc. We have here excluded 2 G and 6 K stars found between 3 and 10pc, but this does not mean that very interesting stars that are more nearby are going to be excluded. Using a longer integration time and more complex observational arrangements also those could be observed. As what concerns M-dwarfs, there are 36 objects between 3pc and 10pc, but on the other hand, these stars tend to have such a small diameter that the leakage will be relatively mild.

In our calculations in Figure 6a-c, we have used a triangular configuration.

The conclusion then is that one could utilize the simpler configurations resulting in a $\theta^2$ null. Less complex architectures that provide a $\theta^2$ null, thus introduce higher stellar leakage and thus longer integration times for close-by stars and at short wavelengths but reduce cost and complexity of implementation. For observations at a wavelength of 10 μm our calculations show a 15% increase of the integration time when observing the Earth around Sun when located at 10pc.

## 8. CONCLUSIONS

The dominating noise source, represented by the local zodiacal cloud, is essentially constant for all target stars while the stellar leakage decreases as the inverse of the distance squared. This means that such stellar leakage has an effect on the integration times of near-by target stars, while for more distant targets its influence decreases significantly. In this study we assessed the impact of different types of nulling profiles, i.e. $\theta^4$ and $\theta^2$, and identify those stars for which the detection sensitivity can be maximized. One has to keep in mind that spectroscopy measurements at the low end of the waveband will require longer integration times. In the case of operating in the mid-infrared as is suggested for ESA's Darwin mission this primarily affect the detection of water at wavelengths larger than 6.5 μm. This line is very important for properly characterizing the atmospheres and physical conditions of Terrestrial Exoplanets, especially as what concerns habitability and when interpreting biomarkers. Other noise sources, such as e.g. what has been termed variability noise impacts also on this region of the spectrum for low-index nulling (Lay & Dubovitsky, 2004; Chazelas et al. – private communication) and need to be addressed accordingly.

The original Darwin mission concept was optimized for stellar rejection to be able to observe also the closest stars for planetary companions. This lead to a baseline concept of a free flying configuration with 6 collector telescopes and a central beam combiner and with a $\theta^4$ shape of the central null. Our analysis (section 3) shows that the starlight rejection criteria can be relaxed while still maintaining a target sample of up to more than 500 stars. Accordingly, candidate Darwin and TPF configurations can use three or four telescopes, reducing complexity and cost of the mission. At the same time, individual telescopes can be made larger and thus extend further out into space. Darwin is in principle (contrary to coronographs) not limited by distance (but by integration time) until the baseline required extends beyond what the transfer optics can handle (due to scattering etc,). Our recent work has demonstrated that Darwin can be implemented with a destructive interference pattern of the $\theta^2$ shape, increasing the stellar leakage but maintaining the science mission goal.